\documentclass[11pt]{article}

\usepackage{amssymb,amsmath,url,epsfig}

\newcommand{\bfa}{{\bf a}}
\newcommand{\bfb}{{\bf b}}
\newcommand{\bfq}{{\bf q}}
\newcommand{\x}{{\bf x}}
\newcommand{\y}{{\bf y}}
\newcommand{\rr}{{\mathbb C}}
\newcommand{\cc}{{\mathbb C}}

\begin{document}

\title{Orthogonalization on a General Purpose Graphics Processing Unit \\
       with Double Double and Quad Double Arithmetic\thanks{This material 
is based upon work supported 
by the National Science Foundation under Grant No.\ 1115777.}}

\author{Jan Verschelde and Genady Yoffe \\
Department of Mathematics, Statistics, and Computer Science \\
University of Illinois at Chicago \\
851 South Morgan (M/C 249) \\
Chicago, IL 60607-7045, USA \\
Emails: \texttt{jan@math.uic.edu} and \texttt{gyoffe2@uic.edu} \\
URLs: \texttt{\url{www.math.uic.edu/~jan}} and
  \texttt{\url{www.math.uic.edu/~gyoffe}}}

\date{13 January 2013}

\maketitle

\begin{abstract}
Our problem is to accurately solve linear systems
on a general purpose graphics processing unit
with double double and quad double arithmetic.
The linear systems originate from the application of Newton's method
on polynomial systems.  
Newton's method is applied as a corrector in a path following method,
so the linear systems are solved in sequence and not simultaneously.
One solution path may require the solution of thousands of linear systems.
In previous work we reported good speedups with our implementation
to evaluate and differentiate polynomial systems on the NVIDIA Tesla C2050.
Although the cost of evaluation and differentiation often dominates 
the cost of linear system solving in Newton's method, 
because of the limited bandwidth of the communication between CPU and GPU,
we cannot afford to send the linear system to the CPU for solving
during path tracking.

\medskip

Because of large degrees, the Jacobian matrix may contain extreme values,
requiring extended precision, leading to a significant overhead.
This overhead of multiprecision arithmetic is our main
motivation to develop a massively parallel algorithm.
To allow overdetermined linear systems we solve linear systems in the
least squares sense, computing the QR decomposition of the matrix by
the modified Gram-Schmidt algorithm.
We describe our implementation of the modified
Gram-Schmidt orthogonalization method for the NVIDIA Tesla C2050,
using double double and quad double arithmetic.  Our experimental
results show that the achieved speedups are sufficiently high to
compensate for the overhead of one extra level of precision.

\medskip

\noindent {\bf Keywords.}
double double arithmetic,
general purpose graphics processing unit,
massively parallel algorithm,
modified Gram-Schmidt method, orthogonalization, 
quad double arithmetic, quality up.

\end{abstract}

\newpage

\section{Introduction}

We consider as given a system of $m$ polynomial equations in $n$ variables.
The coefficients of the polynomials are complex numbers.  Besides~$m$ and~$n$,
two other factors determine the complexity of the system: the number~$M$
of monomials that appear with nonzero coefficient and the largest degree~$d$
that occurs as an exponent of a variable.  The tuple $(m,n,M,d)$ determines
the cost of evaluating and differentiating the system accurately.
As the degrees increase, standard double precision arithmetic becomes
insufficient to solve polynomial systems with homotopy continuation
methods (see for example~\cite{Li03} for an introduction).
In the problem setup for this paper we consider the tracking of
one difficult solution path in extended precision.

The extended precision arithmetic we perform with the
quad double library {\tt QD~2.3.9}~\cite{HLB00},
and in particular on a GPU using the software in~\cite{LHL10}.
For the numerical properties, we refer to~\cite{Dek71} and~\cite{Pri92}.
Our development of massively parallel algorithms is motivated
by the desire to offset the extra cost of double double and
quad double arithmetic.  We strive for a quality up~\cite{Akl04}
factor: if we can afford to keep the execution time constant,
how much can we improve the quality of the solution?

Using double double or quad double arithmetic we obtain predictable
cost overheads.  In~\cite{VY10} we experimentally determined that the
overhead factors of double double over standard double arithmetic is
indeed similar to the overhead of complex over standard double arithmetic.
In terms of quality, the errors are expected to decrease proportionally 
to the increase in the precision.
In~\cite{VY11} we described a multicore implementation of
a path tracker and we implemented 
our methods used to evaluate and differentiate
systems of polynomials were implemented on the NVIDIA Tesla C2050,
as described in~\cite{VY12}.  The focus of this paper is on the
solving of the linear systems, needed to run Newton's method. 

Because of the limited bandwidth of CPU/GPU communication we cannot
afford to transfer the evaluated system and its Jacobian matrix 
from the GPU to the CPU and perform the linear system solving on
the CPU.  Although the evaluation and differentiation of a polynomial
system often dominates the cost of Newton's method~\cite{VY11}, 
the cost of linear system solving increases
relative to the parallel run times of evaluation and differentiation
so that even with minor speedups, using a parallel version of the linear
system solver matters in the overall execution time.

In the next section we state our problem, mention related work
and list our contributions.
In the third section we summarize the mathematical definition
and properties of the modified Gram-Schmidt method and we
illustrate the higher cost of complex multiprecision arithmetic.
Then we describe 
our parallel version of the modified Gram-Schmidt algorithm 
and give computational results.

\section{Problem Statement and Related Work}

Our problem originates from the application of homotopy continuation
methods to solve polynomial systems.  While the tracking of many solution
paths is a pleasingly parallel computation for which message passing
works well, see for example~\cite{SMSW06}, it often occurs that there
is one difficult solution path for which the double precision is
insufficient to reach the solution at the end of the path.
The goal is to offset the extra cost of extended precision
using a GPU.

In this paper we focus on the solving of a linear system
(which may have more equations than unknowns) on a GPU.  
The linear system occurs in the context of
Newton's method applied to a polynomial system.
Because the system could have more equations than unknowns
and because of increased numerical stability, we decided to
solve the linear system with a least squares method via
a QR decomposition of the matrix.  The algorithm we decided
to implement is the modified Gram-Schmidt algorithm,
see~\cite{Hig96} for its definition and a discussion of its
numerical stability.  A computational comparison between Gaussian 
elimination and orthogonal matrix decomposition can be found
in~\cite{TS90}.

Because the overhead factor of the computation cost of extended
precision arithmetic,
we can afford to apply a fine granularity in our parallel algorithm.

\subsection{Related Work}  

Comparing QR with Householder
transformations and with the modified Gram-Schmidt algorithm,
the authors of~\cite{OW90} show that on message passing systems,
a parallel modified Gram-Schmidt algorithm can be much more
efficient than a parallel Householder algorithm, 
and is never slower.  MPI implementations of
three versions of Gram-Schmidt orthonormalizations 
are described in~\cite{Lin00}.  The performance of
different parallel modified Gram-Schmidt algorithms
on clusters is described in~\cite{RS05}.
Because the modified Gram-Schmidt method cannot be expressed by
Level-2 BLAS operations, in~\cite{YTBS06} the authors proposed
an efficient implementation of the classical Gram-Schmidt method.

In~\cite{MS09} is a description
of a parallel QR with classical Gram-Schmidt on GPU and results on
an implementation with the NVIDIA Geforce 295 are reported.
A report on QR decompositions using Householder transformations
on the NVIDIA Tesla C2050 can be found in~\cite{ABDK11}.  
A high performance implementation of the QR algorithm 
on GPUs is described in~\cite{KCR09}.
The authors of~\cite{KCR09} did not consider to implement
the modified Gram-Schmidt method on a GPU because the vectors in the inner
products are large and the many synchronizations incur a prohibitive
overhead.  According to~\cite{KCR09}, a blocked version is
susceptible to precision problems.  In our setting, the length $n$
of the vectors is small (our $n$ may coincide with the warp size)
and similar to what is reported in~\cite{ABDK11}, we expect the cost
of synchronizations to be modest for a small number of threads.
Because of our small dimensions, we do not consider a blocked version.

In~\cite{ASK12}, the problem of solving many small independent 
QR factorizations on a GPU is investigated.
Although our QR factorizations are also small, in our application
of Newton's method in the tracking of one solution path,
the linear systems are not independent and must be solved in sequence.
After the QR decomposition, we solve an upper triangular linear system.
The solving of dense triangular systems on multicore and GPU
accelerators is described in~\cite{TRLD10}.

Triple precision (double + single float) implementations of BLAS
routines on GPUs were presented in~\cite{MT12}.

Related to polynomial system solving on a GPU,
we mention two recent works.  In~\cite{MP11}, a subresultant method
with a CUDA implementation of the FFT is described to solve systems
of two variables.  The implementation with CUDA of
a multidimensional bisection algorithm on an NVIDIA GPU
is presented in~\cite{KP12}.

\subsection{Our Contributions are twofold:} 

\begin{enumerate}
\item We show that the extra 
      cost of multiprecision arithmetic in the modified Gram-Schmidt
      orthogonalization method can be compensated by a GPU.
\item Combined with projected speedups of our massively parallel
      evaluation and differentiation implementation~\cite{VY12},
      the results pave the way for a path tracker that runs
      entirely on a GPU.
\end{enumerate}

\section{Modified Gram-Schmidt Orthogonalization}

Roots of polynomial systems are typically complex
and we calculate with complex numbers.
Following notations in~\cite{GV96},
the inner product of two complex vectors $\x, \y \in \cc^n$
is denoted by $\x^H \y$.  In particular:
$\displaystyle \x^H \y = \sum_{\ell=1}^n \overline{x}_\ell y_\ell$,
where $\overline{c}$ is the complex conjugate of $c \in \cc$.
Figure~\ref{figalgmgs} lists 
pseudo code of the modified Gram-Schmidt orthogonalization method.  
\begin{figure}[hbt]
\begin{center}
\begin{tabbing}
\hspace{8mm} \= Input: $A \in \cc^{m \times n}$. \\
\> Out\=put\=: \= $Q \in \cc^{m \times n}$, $R \in \cc^{n \times n}$: 
                  $Q^H Q = I$, \\
\>    \>   \>  \> $R$ is upper triangular, and $A = QR$. \\
\> let ${\bfa}_k$ be column~$k$ of~$A$ \\
\> for $k$ from 1 to $n$ do \\
\>    \> $r_{kk} := \sqrt{ \bfa_k^H \bfa_k }$ \\
\>    \> $\bfq_k := \bfa_k/r_{kk}$, $\bfq_k$ is column $k$ of~$Q$ \\
\>    \> for $j$ from $k+1$ to $n$ do \\
\>    \>   \> $r_{kj} := \bfq_k^H \bfa_j$ \\
\>    \>   \> $\bfa_j := \bfa_j - r_{kj} \bfq_k$
\end{tabbing}
\caption{The modified Gram-Schmidt orthogonalization algorithm.}
\label{figalgmgs}
\end{center}
\end{figure}

Given the QR decomposition of a matrix~$A$,
the system~$A \x = \bfb$ is equivalent to $QR \x = \bfb$.
By the orthogonality of~$Q$, solving $A \x = \bfb$
is reduced to the upper triangular system $R \x = Q^H \bfb$.
This solution minimizes $|| \bfb - A \x ||_2^2$.

Instead of computing~$Q^H \bfb$ separately,
for numerical stability as recommended in~\cite[\S 19.3]{Hig96},
we apply the modified Gram-Schmidt method to the matrix~$A$
augmented with~$\bfb$:

\begin{equation}
   \left[ 
      \begin{array}{cc}
         A & \bfb
      \end{array}
   \right]
   = 
   \left[
      \begin{array}{cc}
         Q & \bfq_{n+1}
      \end{array}
   \right]
   \left[
      \begin{array}{cc}
         R & \y \\
         0 & z
      \end{array}
   \right].
\end{equation}
As $\bfq_{n+1}$ is orthogonal to the column space of~$Q$,
we have $|| \bfb - A \x ||_2^2 = ||R \x - \y||_2^2 + z^2$
and $\y = Q^H \bfb$.

As reported in~\cite{Hig96},
the number of flops in the algorithm in Figure~\ref{figalgmgs}
equals $2 m n^2$.  In computations we experience the cubic behavior
of the running time: doubling $n$ and $m$ multiplies the running time
by a factor of about~8.


\section{Complex and Multiprecision Arithmetic: Cost and Accuracy}

With user CPU times of runs with the modified Gram-Schmidt algorithm
on random data in
Table~\ref{tabprecision} we illustrate the overhead factor 
of using complex double, complex double double, and
complex quad double arithmetic over standard double arithmetic.
Computations in this section
were done on one core of an 3.47 Ghz Intel Xeon X5690
and with version 2.3.70 of PHCpack~\cite{Ver99}.
Going from double to complex quad double arithmetic,
3.7 seconds increase to 2916.8 seconds (more than 48 minutes),
by a factor of 788.3.

\begin{table}[hbt]
\begin{center}
\caption{User CPU times for 10,000 QR decompositions
         with $n = m = 32$,
         for increasing levels of precision.}
\label{tabprecision}
\begin{tabular}{r|r|r}
             precision &   CPU time & factor  \\ \hline
                double &    3.7 sec & 1.0     \\
        complex double &   26.8 sec & 7.2     \\
 complex double double &  291.5 sec & 78.8    \\
   complex quad double & 2916.8 sec & 788.3 
\end{tabular}
\end{center}
\end{table}

Using the cost factors we can recalibrate the dimension.
Suppose a flop costs 8 times more, using $8 = 2^3$, 
the number of flops in the modified Gram-Schmidt method 
is then $8 \times 2 m n^2 = 2(2m)(2n)^2$.
Working with operations that cost 8 times more increases the cost
with the same factor as doubling the dimension in the original arithmetic.

Taking the cubed roots of the factors in Table~\ref{tabprecision}:
$7.2^{1/3} \approx 1.931$, $78.8^{1/3} \approx 4.287$,
$788.3^{1/3} \approx 9.238$, the cost of using
complex double, complex double double, and complex quad double
arithmetic is equivalent to using double arithmetic, 
after multiplying the dimension 32 of the problem respectively
by the factors 1.931, 4.287, and 9.238, which then yields
respectively 62, 134, and 296.
Orthogonalizing 32 vectors in $\cc^{32}$ in quad double arithmetic 
has the same cost as
orthogonalizing 296 vectors in~$\rr^{296}$ with double precision.

To measure the accuracy of the computed $Q \in \cc^{m \times n}$ 
and $R \in \cc^{n \times n}$ of a given $A \in \cc^{m \times n}$,
we consider the matrix 1-norm~\cite{GV96} of $A - QR$: 
\begin{equation} \label{eqdef1norm}
   e = || A - QR ||_1
     = \max_{\substack{i=1,2,\ldots,m \\ j = 1,2,\ldots,n}}
       \left| a_{ij} - \sum_{\ell=1}^n q_{i\ell} r_{\ell j} \right|.
\end{equation}

For $x \in [-10,+10]$, we have $x^d \in [10^{-d},10^{+d}]$,
so as the degrees $d$ of the polynomials in our system 
increase we are likely to obtain more extreme values 
in the Jacobian matrix.
In the experiments discussed below we generate complex numbers
of modulus one as $\exp(i \theta)$, where $i = \sqrt{-1}$ and
$\theta$ is chosen at random from $[0,2 \pi[$.
To generate complex numbers of varying magnitude,
we consider $r \exp{i \theta}$, with $r$ chosen at random from 
$[10^{-g},10^{+g}]$ where $g$ determines the range of the moduli
of the generated complex numbers.
To simulate the numbers in the Jacobian matrices arising from
evaluating polynomials of degree~$d$, it seems natural to take
the parameter $g$ equal to $d$.

In Table~\ref{tabaccuracy} experimental values for~$e$ are summarized.  
For complex numbers with moduli in $[10^{-g},10^{+g}]$,
$\log_{10}(e)$ decreases linearly as $g$ increases. 
Computing $e$ for 1,000 different random matrices,
$|\min(\log_{10}(e)) - \max(\log_{10})(e))|$
remains almost constant and increases slightly as $g$ increases.  
%

\begin{table}[hbt]
\begin{center}
\caption{For 1,000 QR decompositions on 32-by-32 matrices 
         with randomly generated complex numbers
         uniformly in $[10^{-g},10^{+g}]$, and
         for $g = 1, 4, 8, 12, 16$, we list 
         $m_e = \min(\log_{10}(e))$, $M_e = \max(\log_{10}(e))$,
         and $D_e = m_e - M_e$, computed in complex double and complex
         double double arithmetic.
         For $g = 17, 20, 24, 28, 32$, results are
         for complex double double and complex quad double arithmetic.}
\label{tabaccuracy}
\begin{tabular}{r||r|r|r||r|r|r}
 & \multicolumn{3}{c||}{complex double}
 & \multicolumn{3}{c}{complex double double} \\ 
$g$ & $m_e$ & $M_e$ & $D_e$ & $m_e$ & $M_e$ & $D_e$ \\ \hline
  1 & -14.5 & -14.0 &   0.5 & -30.6 & -30.1 & 0.5 \\
  4 & -11.7 & -11.0 &   0.7 & -27.8 & -27.1 & 0.7 \\
  8 &  -7.8 &  -7.0 &   0.8 & -24.0 & -23.1 & 1.0 \\
 12 &  -3.9 &  -3.1 &   0.8 & -20.1 & -19.2 & 0.9 \\
 16 &  -0.2 &   1.0 &   1.2 & -16.4 & -15.1 & 1.3 \\ \hline \hline
%
 & \multicolumn{3}{c||}{complex double double}
 & \multicolumn{3}{c}{complex quad double} \\ 
$g$ & $m_e$ & $M_e$ & $D_e$ & $m_e$ & $M_e$ & $D_e$ \\ \hline
 17 & -15.5 & -14.1 &  1.3  & -48.1 & -47.1 &  1.0  \\
 20 & -12.6 & -11.1 &  1.5  & -45.1 & -44.2 &  0.9  \\
 24 &  -8.8 &  -7.2 &  1.6  & -41.3 & -40.2 &  1.2  \\
 28 &  -4.7 &  -3.2 &  1.5  & -37.7 & -36.1 &  1.6  \\
 32 &  -1.0 &   0.8 &  1.9  & -33.9 & -32.2 &  1.8
\end{tabular}
\end{center}
\end{table}

Our experiments show the numerical stability of the
modified Gram-Schmidt method to be good and predictable.
If our numbers range in modulus between $10^{-g}$ and $10^{+g}$ and
if we want answers accurate of at least half of our working precision,
then the working precision must be at least $2g$ decimal places.

Our modified Gram-Schmidt method does not swap columns 
(as it must do for rank deficient matrices).
With Gaussian elimination we have to apply partial pivoting
to prevent the growth of the numbers.  
As concluded by~\cite[page~358]{TS90}: ``For QR factorization
with or without pivoting, the average maximum element of the
residual matrix is $O(n^{1/2})$, whereas for Gaussian elimination
it is $O(n)$.''  Even for relatively small dimensions as $n = 32$, 
we have $n/\sqrt{n} \approx 5.66$.  While Gaussian elimination is
3 times faster than the modified Gram-Schmidt method, the average
maximum error is almost 6 times larger.

\section{Massively Parallel Modified Gram-Schmidt Orthogonalization}

Our main kernel {\tt Normalize\_Remove}() in Gram-Schmidt orthogonalization 
normalizes a vector and removes components of all vectors with bigger indexes
in the direction of this vector. 
The secondary kernel {\tt Normalize}() only normalizes one vector.
The algorithm in Figure~\ref{figparalgmgs} overwrites the input matrix~$A$
so that on return the matrix $A$ equals the matrix~$Q$ of the algorithm
in Figure~\ref{figalgmgs}.

\begin{figure}[hbt]
\begin{center}
\begin{tabbing}
\hspace{6mm} \= Input:\= $A \in \cc^{m \times n}$,
                         $A = [ \bfa_1 ~ \bfa_2 ~ \ldots ~ \bfa_n]$, \\
             \>       \> $\bfa_k \in \cc^m$, $k=1,2,\ldots,n$. \\
\> Out\=put\=: \= $A \in \cc^{m \times n}$,
                  $A^H A = I$ (i.e.: $A = Q$), \\
\>    \>   \>  \> $R \in \cc^{n \times n}$: 
                  $R = [ r_{ij} ]$, $r_{ij} \in \cc$, \\
\>    \>   \>  \> $i=1,2,\ldots,n$, $j=1,2,\ldots,n$. \\
\>for $k$ from 1 to $n-1$ do \\
\>\>  launch kernel {\tt Normalize\_Remove}($k$)\\
\>\> with $(n-k)$ blocks of threads, \\
\>\> as the $j$th  block (for all $j: k < j \leq n$) \\
\>\>  normalizes $\bfa_k$ and removes the component \\
\>\> of $\bfa_j$ in the direction of $\bfa_k$ \\
\> launch kernel {\tt Normalize}($n$) with one \\
\> thread block to normalize $\bfa_n$. 

\end{tabbing}
\caption{A parallel version of the modified Gram-Schmidt 
         orthogonalization algorithm.}
\label{figparalgmgs}
\end{center}
\end{figure}

The multiple blocks launched by the kernel within each iteration 
of the loop in the algorithm in Figure~\ref{figparalgmgs} 
is the first coarse grained level of parallelism.
If the number of variables equals the warp size
(the number of cores on a multiprocessor of the GPU),
then the second fine grained parallelism resides in the calculation
of componentwise operations and of the inner products. 
Threads within blocks perform these operations cooperatively.
As one inner product of two vectors of dimension~$n$
requires $n$ multiplications (one operation per core),
note that a multiplication in double double and quad double arithmetic 
requires many operations with hardware doubles.

The algorithm suggests the normalization of each $\bfa_k$
is performed $(n-k)$ times, by each of the blocks in the $k$th launch 
of the kernel {\tt Normalize\_Remove}(). 
However normalizing it only once instead would suggest another 
launch of the kernel {\tt Normalize}() associated with extra writing 
to and reading from  the global memory of the card of the vector 
being normalized. 
This would be more expensive than to perform the normalization 
within {\tt Normalize\_Remove}() multiple times.

The loop in the algorithm in Figure~\ref{figparalgmgs} performs
$n-1$ normalizations, where each normalization is followed by
the update of all remaining vectors.
In particular, after normalizing the $k$th vector, we launch $n-k$ blocks
of $m$ threads.  Each thread block handles one $\bfa_j$.
The update stage has a triangular structure.
The triangular structure implies that we have more parallelism
for small values of~$k$.
Therefore, we expect increased speedups at earlier stages
of the algorithm in Figure~\ref{figparalgmgs}.

The main ingredients in the kernels {\tt Normalize}() and
{\tt Normalize\_Remove}() are inner products and the normalizations,
which we explain in the next two subsections.
In subsection~C we discuss the usage of the card resources by
threads of the kernel {\tt Normalize\_Remove}().

\subsection{Computing Inner Products}

The fine granularity of our parallel algorithm is explained in this section.
In computing $\x^H \y$ the products $\overline{x}_{\ell} \star y_{\ell}$
are independent of each other.  
The inner product $\x^H \y$ is computed in two stages:
\begin{enumerate}
\item All threads work independently in parallel:
      thread $\ell$ calculates $\overline{x}_{\ell} \star y_{\ell}$
      where the operation $\star$ is a complex double,
      a complex double double, or a complex quad double multiplication.
      Afterwards, all threads in the block are synchronized
      for the next stage.
\item The application of a sum reduction~\cite[Figure~6.2]{KH10} 
      to sum the elements in 
      $(\overline{x}_1 y_1, \overline{x}_2 y_2, \ldots, \overline{x}_m y_m)$
      and compute
      $\overline{x}_1 y_1 + \overline{x}_2 y_2 + \cdots + \overline{x}_m y_m$.
      The $+$ in the sum above corresponds to the $\star$ in the item above
      and is a complex double, a complex double double, or a complex quad 
      double addition.  There are $\log_2(m)$ steps but as the $m$ typically
      equals the warp size, there is thread divergence in every step.
\end{enumerate}
The number of shared memory locations used by an inner product equals~$2 m$.
Each location holds a complex double, or a complex double double,
or a complex quad double.

The $2 m$ memory locations suffice if we compute 
only one inner product, allowing
that one of the original vectors is overwritten.
In our algorithm, we need the same vector $\bfq_k$ the second time
when computing $r_{kj} := \bfq_k^H \bfa_j$ (see Figure~\ref{figalgmgs})
so we need an extra $m$ shared memory locations to store 
$\overline{q}_{k \ell} \star a_{j \ell}$ for $\ell = 1,2,\ldots,m$.
Storing $r_{kj}$ in a register, the extra $m$ shared memory locations
are reused to store the products
$r_{kj} \star q_{k \ell}$ for $\ell = 1,2,\ldots,m$,
in the computation of $\bfa_j := \bfa_j - r_{kj} \bfq_k$.
So in total we have $3m$ memory locations in shared memory in
the kernel {\tt Normalize\_Remove}().

\subsection{The Orthonormalization Stage}

After the computation of the inner product $\bfa_k^H \bfa_k$,
the orthonormalization stage consists of
one square root computation, followed by $m$ division operations.

The first thread in a block performs the square root calculation 
$r_{kk} := \sqrt{\bfa_k^H \bfa_k}$ and then, after a synchronization,
the $m$ threads in a block independently perform in-place divisions
$a_{k \ell} := a_{k \ell}/r_{kk}$, for $\ell =1,2,\ldots,m$ to 
compute~$\bfq_k$.

Dividing each component of a vector by the norm happens independently, 
and as the cost of the division increases for complex doubles, complex 
double doubles, and complex quad doubles, we could expect an
increased parallelism as we increase the working precision.
Unfortunately, also the cost for the square root calculation
--- executed in isolation by the first thread in each block ---
also increases as the precision increases.

\subsection{The Occupancy Of Multiprocessors}

Using the CUDA GPU Occupancy Calculator, we take $m = n$,
and consider the use of complex double double arithmetic.
Concerning the occupancy of the multiprocessors,
the $3m$ vectors in one thread block take
\begin{equation}
   3 \times n \times \mbox{size\_of(complex double double)}
\end{equation}
bytes of shared memory.  For $n = 32$, 
this amounts to $3 \times 32 \times 32 = 3,072$ bytes. 
Also a thread block uses $48 \times 32 = 1,536$ registers. 
The number of blocks scheduled per multiprocessor is~8.
It is actually the maximum number of blocks which could be scheduled per
multiprocessor for the device of compute capability 2.0.
Allocated per block shared memory, 
and the number of registers used 
do not appear as the limiting factor on the number
of blocks scheduled per multiprocessor.
Although shared memory and
registers of a multiprocessor are employed quite well:
8 blocks of threads use about 
(we multiply by 100 to get a percentage)
\begin{equation}
   100 \times 3,136 \times 8/49,152 \approx 51\%
\end{equation}
of shared memory capacity, and 
\begin{equation}
  100 \times 1,536 \times 8/32,768 \approx 38\%
\end{equation}
of available registers.
For dimension 32, the orthogonalization launches the kernel 
{\tt Normalize\_Remove()} 31 times,
while first 7 of these launches employ 4 multiprocessors,
launches from 8 to 15 employ 3 multiprocessors,
16-23 employ 2 multiprocessors, 
and finally launches 24-31 employ only one multiprocessor.

\subsection{Data Movement}

At the beginning of the kernel thread $\ell$ of a block 
reads the $\ell$th component of the vector $\bfa_k$ from the global memory 
into the $\ell$th location of the first column of the shared memory 
3-by-$m$ array {\tt Sh\_Locations}
of complex numbers of the given precision allocated by the block.
Subsequently it reads the $\ell$th component of the vector $\bfa_j$ into the
$\ell$th location of the second column of {\tt Sh\_Locations}.
Both readings are done simultaneously by threads of the block.

For double and double double precision levels we have achieved
coalesced access to the global memory but we did not achieve
coalesced access for complex quad double numbers.
This could explain why the speedups do not increase 
as we go from complex double double to the complex
quad double versions of the parallel Gram-Schmidt algorithm.
We think that by reorganizing the storage of complex quad doubles
we can also achieve coalesced memory access 
for arrays of complex quad doubles.

\section{The Back Substitution Kernel}

After the computation of $Q$ and $R$, denoting $Q^H \bfb$ by $\y$,
we have to solve the triangular system $R \x = \y$.
Because of the low dimension of our application,
only one block of threads will be launched.
Pseudo code for a parallel version of the back substitution
algorithm is shown in Figure~\ref{figalgbacksubs}.

The natural order for the parallel version of the back substitution
is to process the matrix $R$ in a column fashion.
In the $k$th step we multiply the $k$th column of $R$ by $x_k$
and subtract the product from the right hand side vector~$\y$
updated by such subtractions at all the previous steps.

\begin{figure}[hbt]
\begin{center}
\begin{tabbing}
\hspace{8mm} \= Inp\=ut: \= $R \in \cc^{n \times n}$,
                          an upper triangular matrix, \\
             \>    \>    \> $\y \in \cc^n$, the right hand side vector. \\
\> Output: $\x$ is the solution of $R \x = \y$. \\
\> for $k$ from $n$ down to 1 do \\
\> \> thread $k$ does $x_k := y_k/r_{kk}$ \\
\> \> for $j$ from 1 to $k-1$ do \\
\> \>  \> thread $j$ does $y_j := y_j - r_{jk} \star x_k$ \\
\end{tabbing}
\caption{Pseudo code for a parallel back substitution.}
\label{figalgbacksubs}
\end{center}
\end{figure}

Ignoring the cost of synchronization and thread divergence
and with the warp size equal to the dimension~$n$,
the parallel execution reduces the inner loop to one step.
With focus on the arithmetical cost, the total number of steps
equals~$2n$.  Note that the more costly division operator is
done by only one thread.  More precisely than $2n$, the
arithmetical cost of the algorithm in Figure~\ref{figalgbacksubs}
is $n$ divisions, followed by $n$ multiplications 
and $n$ subtractions.

During the execution of the parallel back substitution,
the right hand side vector $\y$ remains in shared memory.
At each step $k$, the current column~$k$ of $R$ is loaded
into shared memory for processing.


\section{Massively Parallel Path Tracking}

Typically along a path we solve thousands of linear systems in the
application of Newton's method as the corrector.
The definition of the polynomial system (its coefficients and degrees)
is transferred only once from the host to the card.
The evaluated Jacobian matrix and system never leave the card.
What goes back to the host are the computed solutions along the path.

We close with some observations comparing polynomial evaluation and
differentiation with that of linear system solving.
Based on the experience with our algorithms of~\cite{VY12},
there appears to be less data parallelism in the modified Gram-Schmidt
than with polynomial evaluation and differentiation.
On the other hand, the cost of evaluating and differentiating polynomials
depends on their size, whereas the cost for general linear system solving
remains cubic in the dimension.  For very sparse polynomial systems,
the cost for solving linear systems will dominate.

\section{Computational Results}

Our computations were done on an HP Z800 workstation running Red Hat
Enterprise Linux.  For speedups, we compare the sequential run times
on one core of an 3.47 Ghz Intel Xeon X5690.  The clock speed of the
NVIDIA Tesla C2050 is at 1147 Mhz, about three times slower than the
clock speed of the CPU.  The C++ code is compiled with version 4.4.6
of gcc and we use release 4.0 of the NVIDIA CUDA compiler driver.

\begin{figure}[hbt]
\begin{center}
\epsfig{figure=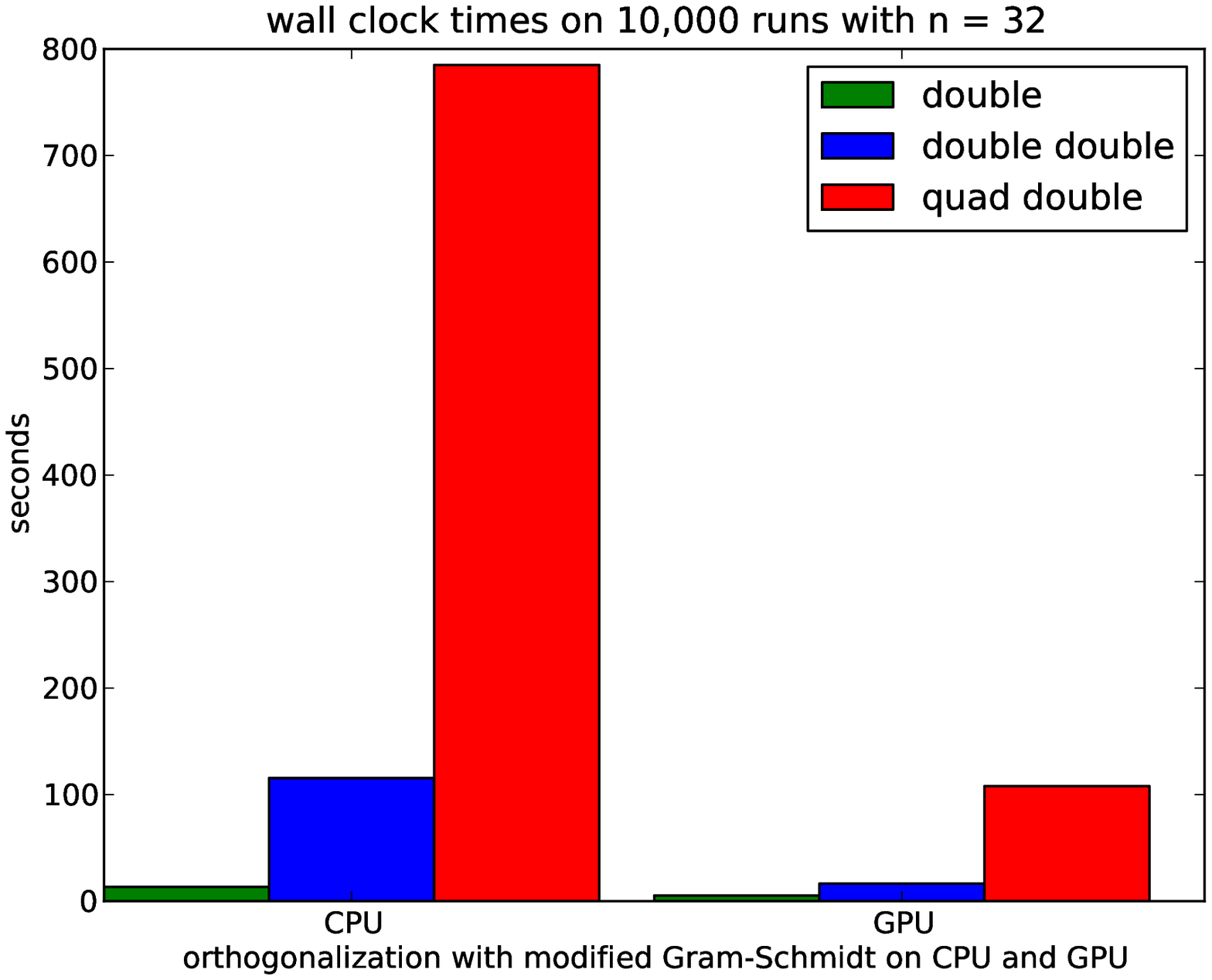,width=9cm}
\end{center}
\caption{Plot corresponding to the data in Table~\ref{tabresults}.
         Observe that the rightmost bar representing times on a GPU
         with quad double arithmetic is shorter than the corresponding
         middle bar on a CPU with double double arithmetic.
         This plot illustrates the compensation of the overhead of
         quad double arithmetic (versus double double arithmetic) by a GPU. }
\label{figtab6barplot}
\end{figure}

We ran the modified Gram-Schmidt method on 32 random complex vectors
of dimension~32, as 32 is the warp size of the NVIDIA Tesla C2050.
The times and speedups are shown in Table~\ref{tabresults}.
The times of Table~\ref{tabresults} are the heights of the bars in
Figure~\ref{figtab6barplot}.

\begin{table}[hbt]
\begin{center}
\caption{Wall clock times and speedups for 10,000 orthogonalizations
         on 32 random complex vectors of dimension 32.}
\label{tabresults}
\begin{tabular}{c|r|r|r}
            precision & 1 CPU core & Tesla C2050 & speedup \\ \hline
       complex double &  13.4~sec &   5.3~sec & 2.5~~ \\
complex double double & 115.6~sec &  16.5~sec & 7.0~~ \\
  complex quad double & 785.0~sec & 108.0~sec & 7.3~~ 
\end{tabular}
\end{center}
\end{table}

The small speedup of 2.5 for complex double arithmetic 
in Table~\ref{tabresults} shows
that the fine granularity of the orthogonalization pays off only 
with multiprecision arithmetic.
Comparing time on the GPU with quad double arithmetic to the
time on the CPU with double double arithmetic we observe that
the 115.6 seconds on one CPU core is of the same magnitude as
the 108.0 seconds on the GPU.  Obtaining more accurate orthogonalizations
in about the same time is quality up.

\begin{figure}[hbt]
\begin{center}
\epsfig{figure=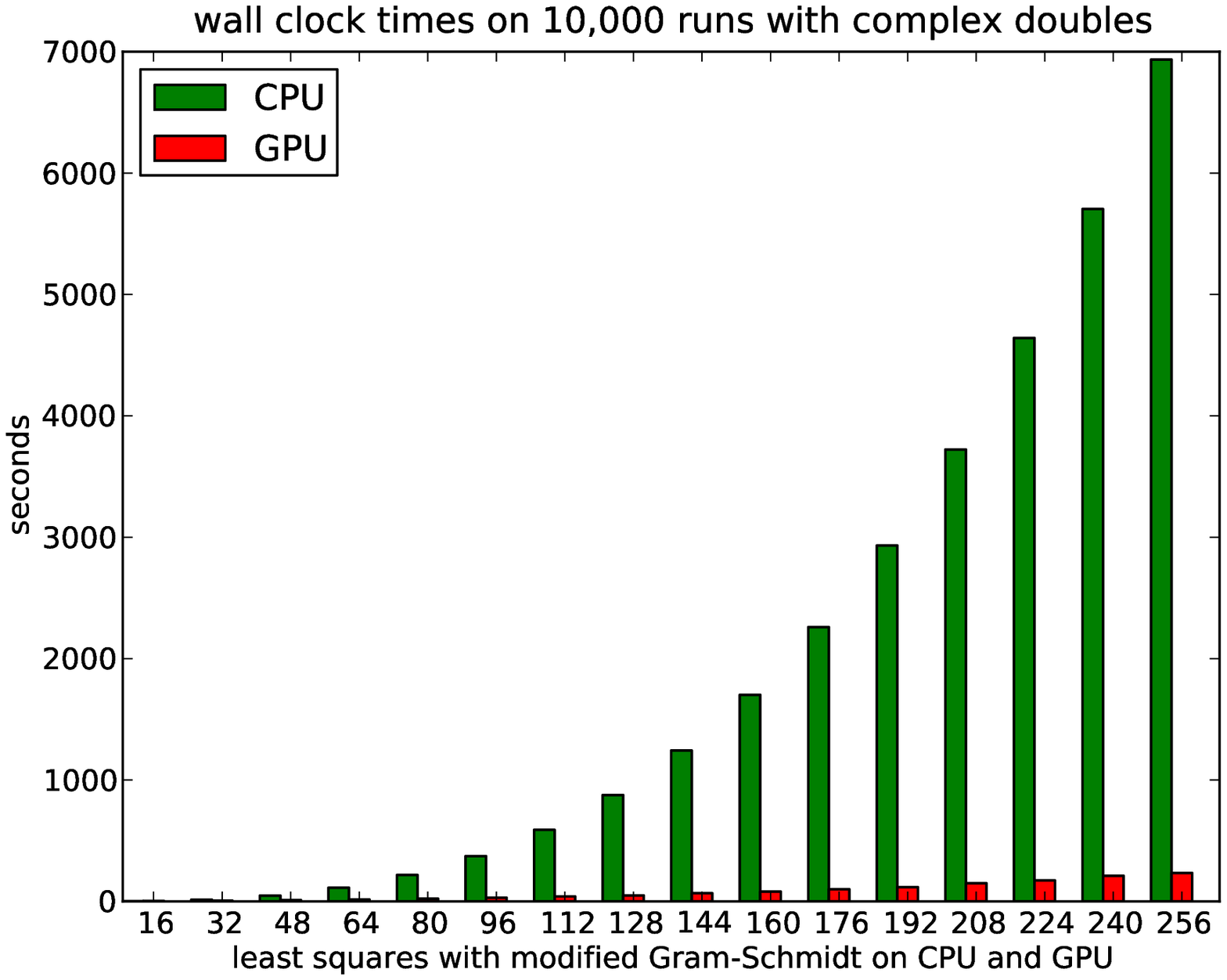,width=9cm}
\end{center}
\caption{Plot corresponding to the data in Table~\ref{tabDmgs},
         for dimensions ranging from $n = 16$ to 256,
         with increments of 16. }
\label{figtabDbarplot}
\end{figure}

We obtain double digit speedups
with complex double arithmetic for $n \geq 96$,
see Table~\ref{tabDmgs} and Figure~\ref{figtabDbarplot}.

\begin{table}[hbt]
\begin{center}
\caption{Wall clock times for 10,000 runs 
         of the modified Gram-Schmidt method
        (each followed by one backsubstitution)
        in complex double arithmetic
        for various dimensions n on CPU and GPU,
        with corresponding speedups.}
\label{tabDmgs}
\begin{tabular}{r||r|r|r}
\multicolumn{4}{c}{complex double arithmetic} \\
 $n$ &    CPU  &    GPU & speedup \\ \hline
 16  &    2.01 &   4.11 &  0.49~ \\
 32  &   14.61 &   6.52 &  2.24~ \\
 48  &   47.80 &  11.11 &  4.30~ \\
 64  &  112.60 &  15.38 &  7.32~ \\
 80  &  217.52 &  22.89 &  9.50~ \\
 96  &  373.06 &  30.43 & 12.26~ \\
112  &  589.35 &  40.82 & 14.44~ \\
128  &  876.11 &  49.10 & 17.84~ \\
144  & 1243.26 &  67.41 & 18.44~ \\
160  & 1701.57 &  80.42 & 21.16~ \\
176  & 2260.07 &  99.94 & 22.61~ \\
192  & 2932.15 & 116.90 & 25.08~ \\
208  & 3722.77 & 149.45 & 24.91~ \\
224  & 4641.71 & 172.30 & 26.94~ \\
240  & 5703.77 & 211.30 & 26.99~ \\
256  & 6935.10 & 234.29 & 29.60~ 
\end{tabular}
\end{center}
\end{table}

\begin{figure}[hbt]
\begin{center}
\epsfig{figure=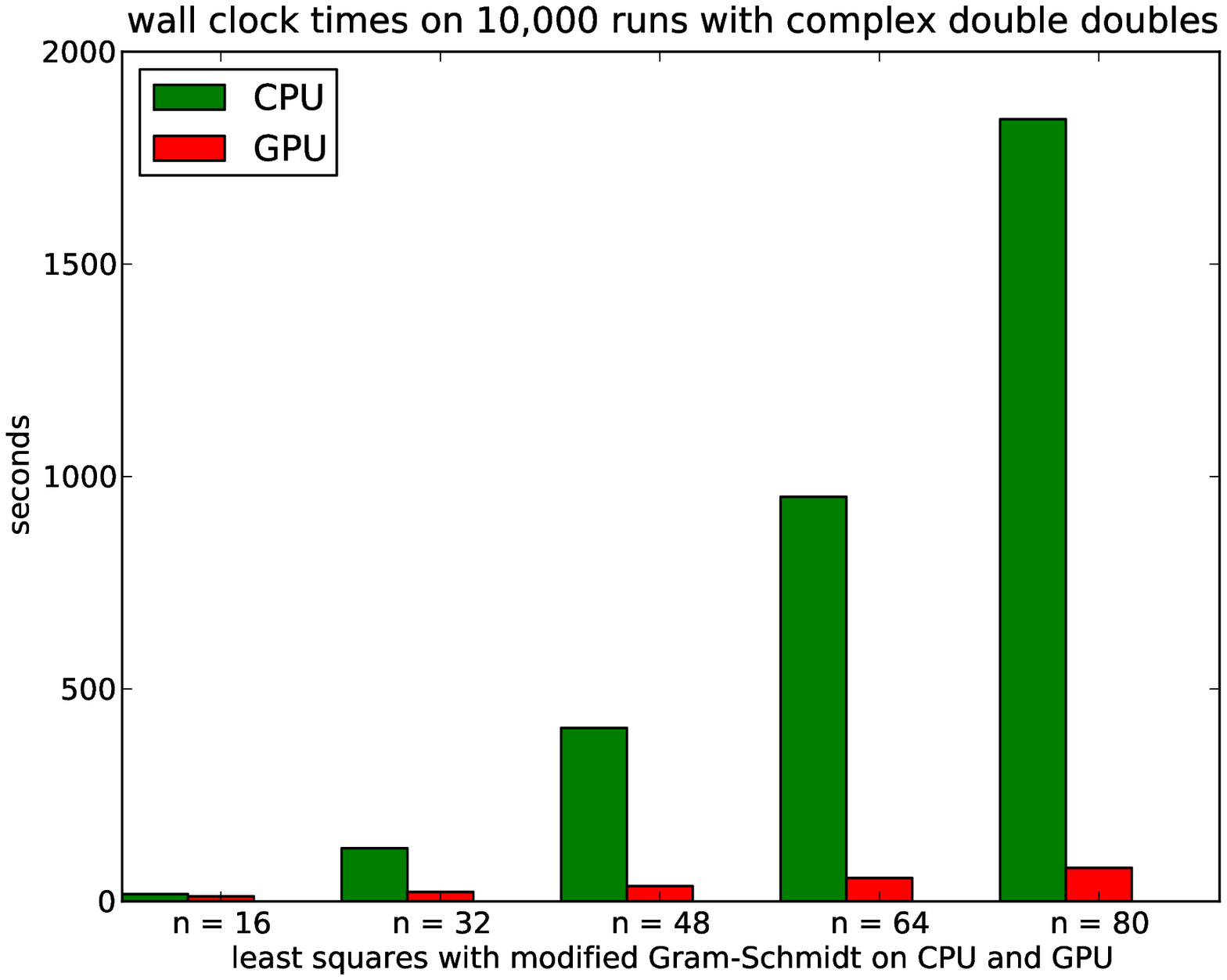,width=9cm}
\epsfig{figure=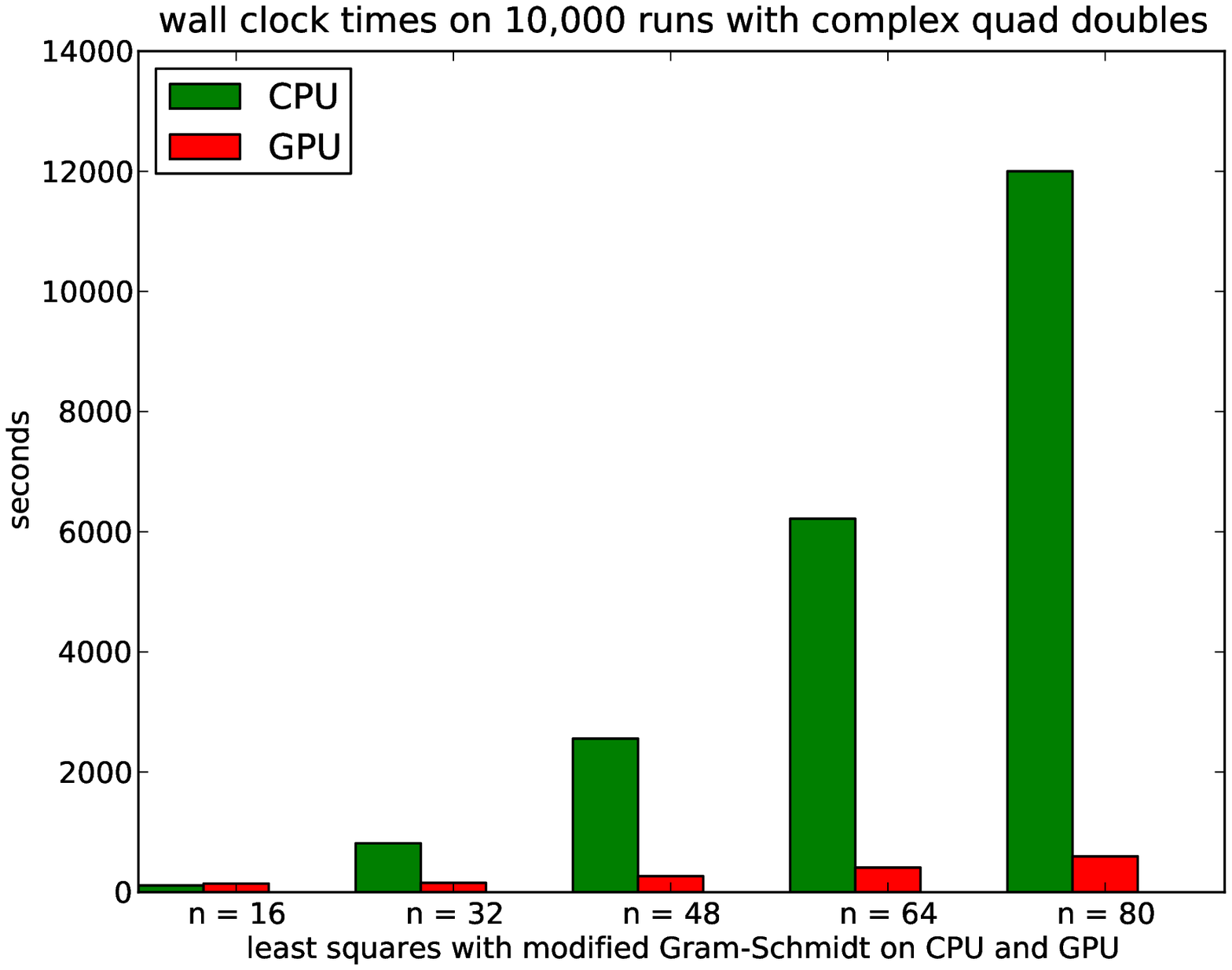,width=9cm}
\end{center}
\caption{Plots corresponding to the data in Table~\ref{tabDDmgs},
         for dimensions $n = 16, 32, 48, 64, 80$. }
\label{figtabDDbarplot}
\end{figure}

Double digits speedups
with complex double double arithmetic 
are obtained for $n \geq 48$.
For complex quad double arithmetic 
we obtain a speedup of almost~10 for $n = 48$.
See Table~\ref{tabDDmgs} and Figure~\ref{figtabDDbarplot}.

Comparing the first and the last execution time on the last line
of Table~\ref{tabDDmgs} we observe the quality up.
For $n = 80$, the GPU computes the least squares solution 
that is twice as accurate in complex quad doubles as the
solution in complex double doubles on the CPU,
computed three times faster than the CPU,
as $1841.07/597.47 = 3.08$.

\begin{table}[hbt]
\begin{center}
\caption{Wall clock times for 10,000 runs 
of the modified Gram-Schmidt method
(each followed by one backsubstitution)
in complex double double and complex quad double arithmetic,
for various dimensions n on CPU and GPU,
with corresponding speedups.}
\label{tabDDmgs}
\begin{tabular}{r||r|r|r||r|r|r}
& \multicolumn{3}{c||}{complex double double} 
& \multicolumn{3}{c}{complex quad double} \\
$n$ &    CPU  &    GPU & speedup &    CPU   &   GPU  & speedup \\ \hline
 16 &   17.17 &  11.85 &  1.45~  &   113.51 & 143.07 &  0.79~ \\
 32 &  125.06 &  22.44 &  5.57~  &   813.65 & 155.32 &  5.24~ \\
 48 &  408.20 &  35.88 & 11.38~  &  2556.36 & 266.55 &  9.59~ \\
 64 &  952.35 &  55.18 & 17.26~  &  6216.06 & 409.57 & 15.18~ \\
 80 & 1841.07 &  79.11 & 23.27~  & 12000.15 & 597.47 & 20.08~
\end{tabular}
\end{center}
\end{table}

\begin{figure}[hbt]
\begin{center}
\epsfig{figure=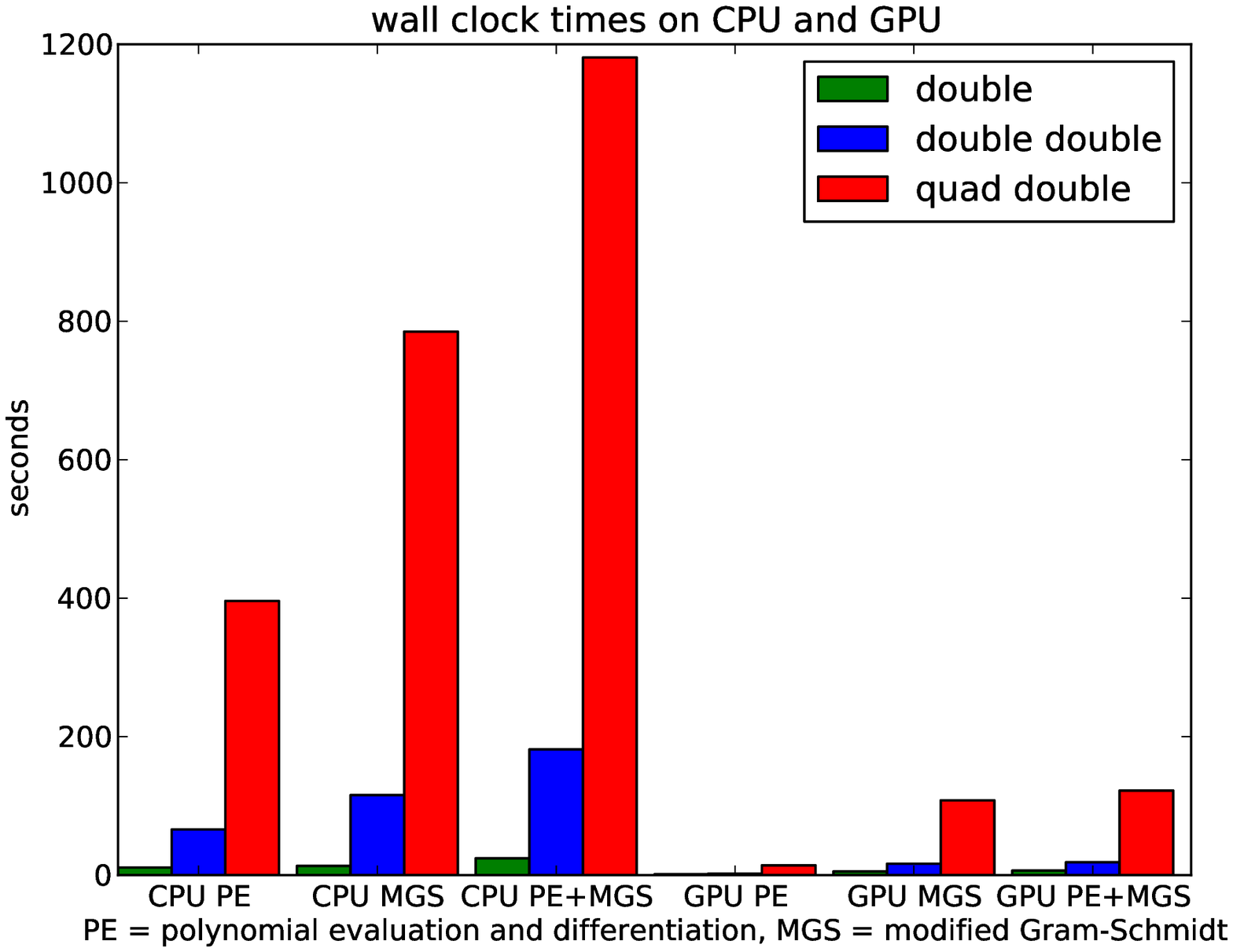,width=9cm}
\end{center}
\caption{Plot corresponding to the data in Table~\ref{PolSys55}.
         Observe that the rightmost bars representing times on a GPU
         with quad double arithmetic are shorter than the corresponding
         middle bars on a CPU with double double arithmetic.
         This plot illustrates the compensation of the overhead of
         quad double arithmetic (versus double double arithmetic) by
         a massively parallel algorithm on a GPU.  }
\label{figtab7barplot}
\end{figure}

We end this paper with one sample of timings to illustrate 
the relationship with polynomial evaluation and differentiation
(applying our parallel algorithms of~\cite{VY12}),
see Figure~\ref{figtab7barplot}. 
The total speedup is still sufficient to compensate for one level
of extra precision.

\begin{table}[hbt]
\begin{center}
\caption{Wall clock times and overall speedups for 10,000 orthogonalizations
         on 32 random complex vectors of dimension 32 and for 10,000 
         polynomial evaluations and differentiations
         of  polynomial system of 32 equations with 32 variables, 
         with 32 monomials per polynomial, with 5 variables in
         each monomial, with variable degrees uniformly taken
         from $\{ 1,2,3,4,5 \}$. }
\label{PolSys55}
\begin{tabular}{c|r|r|r}
            precision &    CPU PE  &     GPU PE & speedup \\ \hline
       complex double &   11.0~sec &    1.3~sec &  8.5~~ \\
complex double double &   66.0~sec &    2.1~sec & 31.4~~ \\
  complex quad double &  396.0~sec &   14.2~sec & 27.9~~ \\ \hline
            precision &    CPU MGS &    GPU MGS & speedup\\ \hline
       complex double &   13.4~sec &    5.3~sec &  2.5~~ \\
complex double double &  115.6~sec &   16.5~sec &  7.0~~ \\
  complex quad double &  785.0~sec &  108.0~sec &  7.0~~ \\ \hline
            precision & CPU PE+MGS & GPU PE+MGS & speedup\\ \hline
       complex double &   24.4~sec &    6.6~sec &  3.7~~ \\
complex double double &  181.6~sec &   18.6~sec &  9.8~~ \\
  complex quad double & 1181.0~sec &  122.2~sec &  9.7~~ 
\end{tabular}
\end{center}
\end{table}

\section{Conclusions}

Using a massively parallel algorithm for the modified Gram-Schmidt
orthogonalization on a NVIDIA Tesla C2050 Computing Processor
we can compensate for the cost of one extra level of precision,
even for modest dimensions, using a fine granularity.
For larger dimensions we obtain double digit speedups
and the GPU computes solutions, twice as accurate
faster than the~CPU.






%


\end{document}